\documentstyle[12pt]{article}
\begin{document}
\begin{center}

{\huge{{\bf Detecting $\nu_{\tau}$ Oscillations at $PeV$ Energies}}}\\

\bigskip

\small{\rm{John~G.~Learned and Sandip~Pakvasa}}\\

\small{{\it Department of Physics and Astronomy, University of Hawaii}}\\
\small{{\it Honolulu, Hawaii 96822 USA}}\\

\bigskip
\bigskip

\small{\it{Preprint UH-511-799-94}}\\
\small{\it{hep-ph/9408296}\\
\small{\it{Submitted to Astroparticle Physics}}\\
\small{\it{15 August 1994}}\\

\bigskip
\bigskip

Abstract
\end{center}

It is suggested that a large deep underocean (or ice) neutrino detector, given
the presence of significant numbers of neutrinos in the PeV energy range as
predicted by various models of Active Galactic Nuclei, can make unique
measurements of the properties of neutrinos.  It will be possible to observe
the existence of the $\nu_{\tau}$, measure its mixing with other flavors, in
fact test the mixing pattern for all three flavors based upon the mixing
parameters suggested by the atmospheric and solar neutrino data, and measure
the $\nu_{\tau}$ cross section.  The key signature is the charged current
$\nu_{\tau}$ interaction, which produces a double cascade, one at either end of
a lightly radiating track.  At a few PeV these cascades would be separated by
roughly $100\ m$, and thus be easily resolvable in next generation DUMAND-like
detectors.  First examples might be found in detectors presently under
construction.  Future applications are precise neutrino astronomy and earth
tomography.

\newpage

\section{The Double Bang Signal}

The interaction of high energy tau neutrinos ($\nu_{\tau}$'s) in DUMAND-like
detectors \cite{DUMAND} will present a spectacular ``double bang''  signature.
The existence of such events depends upon the presence of $10^{15}\ eV$
neutrinos in  adequate numbers, as are in fact predicted from active galactic
nuclei (AGN)\cite{Stenger92} for example.  The interesting signals are from
the charged current ($cc$) quark interactions of $\nu_{\tau}$'s.  Since the
$\tau$ mass is about $1.8\ GeV$, a $\tau$ of $1.8\ PeV$ and with $c \tau$ of
$91\ \mu m$\cite{PDG92} would fly roughly $90\ m$ before decay.   The
signature, as illustrated in Figure \ref{fig:DOUBLEBANG}, is:

\begin{enumerate}

\item    a big hadronic shower from the initial $\nu_{\tau}$ interaction,

\item    a muon like $\tau$ track,  and then

\item    a second and even larger particle cascade.

\end{enumerate}

To give some scale to this, the ratios of detectable photons from these three
segments are roughly $10^{11}$ : $2\cdot 10^6$ : $2\cdot 10^{11}$.  Such large
bursts of light would be visible from distances of hundreds of meters by
present technology photomultipliers.  The charged $\tau$ will be hard to
resolve from the bright Cherenkov light of the cascades, and the photon arrival
times will not be very different.  However, simply connecting the two cascades
by the speed of light will suffice to make an unambiguous association
(including direction of the cascades) of the two bursts.   In fact, as we
discuss later, it appears that the double bang signature alone is nearly
background free, sufficiently so as not to have to invoke lack of either
incoming or outgoing charged particles (muons) to produce a clean sample.  The
double bang event topology appears to be a unique signal for real $\tau$
production by $\nu_{\tau}$'s, thus permitting ``discovery'' of the
$\nu_{\tau}$, and inferring mixing of neutrino flavors. Finding even one of
these events would have significant implications.

The rate for such events is of course unknown experimentally now. If we employ
optimistic fluxes from the Szabo and Protheroe model\cite{Szabo94} for the sum
of neutrinos from  all AGN, we estimate 1000 events of this type per year in a
volume of $1~km^3$ of water or ice, in a $2~PeV$ energy band.  While the
experiments now being constructed (AMANDA, Baikal, DUMAND, and NESTOR) would
expect to see about one such event per year; they should easily determine if
the AGN flux is present and if lucky may find exemplary double bang events.

We show below that $\nu_{\tau}$'s are unlikely to originate in commonly
considered astrophysical sources, but are likely to appear due to neutrino mass
and mixing, over a large range of allowable (and even favored, if the solar
neutrino puzzle and the atmospheric neutrino anomaly have anything to do with
oscillations) neutrino mixing parameter space. The point is that in the general
energy range of a few $PeV$ there exists a powerful tool for searching for
$\tau$ mixing, over an unequalled parameter space, with unambiguous
identification of the $\tau$. We know of no other way to make a $\nu_{\tau}$
{\bf appearance} experiment with cosmic rays, no way has been proposed for an
accelerator experiment except for the use of emulsions making observations of
relatively large $\delta m^2 > 1~eV^2$, and no way of detecting $\nu_{\tau}$'s
except statistically at proposed long baseline accelerator experiments.

In the following we explore the physics implications of the observations of the
double bang events in a little more detail, discussing the kinematics, sources
of $\nu_{\tau}$'s, the sensitivity to two and three neutrino mixing, and
potential (and we conclude small) backgrounds.

\section{Essentially Full Kinematics from Double Bang Events}

One may in principle measure the total energy of the incident neutrino and
nearly the full kinematics of the double bang events by adequate sampling of
the Cherenkov radiation. The observation of light from the first cascade yields
the energy transferred to the quark, $E_1$. (The fraction of energy transfered
to the quark is traditionally designated as $y$).  The magnitude of the light
from the tau track plus the flash of light from the second cascade gives the
energy ($E_2$) kept by the $\tau$ except for that carried away by the decay
neutrino(s).  The sum ($E_T = E_1 + E_2$) gives the rough incoming neutrino
energy, and the ratio of the first cascade energy to the total energy
($E_1/E_T$) provides the $\sim y$ value.  The cross sections and $<y>$ are
almost equal for $\nu_{\tau}$ and $\bar{\nu}_{\tau}$ at this
energy\cite{Quigg86}.  Observing the $y$ distribution is a check on the
observations, and departures from expectations could signal new physics. In
calculating the $\nu_{\tau}$ flux, the measured $y$ distribution will permit
correction for the potentially unobserved events near $y = 0$ (no initial
cascade) and near $y = 1$ (initial cascade with most of the energy and the tau
decays too close to the first cascade for resolution).  The near equality of
the cross sections for particle and antiparticle permits the total flux to be
calculated independently of the mix in the cosmic beam.

In a $\nu_\tau \ cc$ interaction, $<y>$ is about 0.25 at these  energies [4]
and the energy deposited is $<E_1> \sim 1/4\  E_\nu$. The $\tau$ carries about
0.75 times the $\nu_\tau$ energy.   In the subsequent decay of $\tau$, the
energy deposited will  be about $2/3 \ E_\tau \approx 1/2\ E_\nu$ when the
decay
is  hadronic (which is about 64\% of the time).  Hence the  ratio of the
average energy of the second ``bang" to the  first one is given by $E_2/E_1
\sim \ 2$.  This is  characteristic of these $\nu_\tau$ events and making a cut
of $E_2/E_1 \ > 1$ eliminates several kinds of background events as we  discuss
later (see Figure \ref{fig:e1_vs_e2}).  Detailed calculations for  signal as
well as background $E_1$ versus $E_2$ distribution will be presented elsewhere.

The threshold energy for discriminating two cascades will be determined by
requiring a $\tau$ that flies far enough so that the two cascades can be
distinguished, and so that there are no ``punch through'' events.  This
distance will be of the order of some few times the cascade length (order $10\
m$), and thus our threshold for $\tau$ detection via this means would be, as
suggested above, about a $PeV$.   Aside from the physics limitation of several
tens of meters, there will also be a detector dependent limitation depending
upon detector density and response, but probably of similar magnitude.

We note that the observation of the double bang events presents the opportunity
to measure the $PeV$ $\nu_{\tau}$ cross section via the angular distribution in
the lower hemisphere which decreases towards the nadir due to attenuation
through the earth ($\sim 90\%$ for straight upwards travelling neutrinos in
this energy range).  For future studies of earth tomography, the potential of
this process is great, since it does not depend upon convolution over the $y$
distribution and muon range, as is necessary to extract information from the
upcoming muon flux alone.

Also, given the enormous light output of the cascades one would expect that
timing from the detectors (at intermediate distances, since nearby detectors of
present design will surely be saturated) would give excellent vertex
resolution, and thus the initial neutrino direction to a precision of the order
of $1^{\circ}$.  In principle, of course, having both cascades and almost all
energy ``visible'' one can deduce the initial neutrino direction with arbitrary
precision, perhaps making optical precision ultimately possible in some future
neutrino telescope.

\section{Deducing the Neutrino Flavor Mix}

{}From the ensemble of measurements with a DUMAND-like array we will have:

\begin{enumerate}

\item 	The $\tau$ rate from double bang events gives the $\nu_{\tau} +
\bar{\nu}_{\tau}$ flux.

\item 	Measurents of the UHE muon flux permits calculation the $\nu_{\mu} +
\bar{\nu}_{\mu}$ flux.

\item 	The $W^-$ resonant event rate yields the $\bar{\nu}_e$ flux at
$6.4\ PeV$.

\item 	Observation of the cascade rate (as a function of energy) produces the
sum of neutral current interactions of all flavors of neutrinos, and charged
currents without visible $\mu$'s and $\tau$'s (mostly $\nu_e$'s).

\end{enumerate}

If we write $r\ \equiv\ \sigma_{CC} / \sigma_{NC}$, the ratio of charged to
neutral current cross sections, and we note that the cross sections are nearly
flavor and charge independent at this energy, we can summarize the four
observations above as:

\begin{equation}
N_1 = N_{\tau} + N_{\bar{\tau}}
\end{equation}

\begin{equation}
N_2 = N_{\mu} + N_{\bar{\mu}}
\end{equation}

\begin{equation}
N_3 = N_{\bar{e}}
\end{equation}

\begin{equation}
N_4 = (N_{e} + N_{\bar{e}})\cdot (1 + r) +
(N_{\mu} + N_{\bar{\mu}}) +
(N_{\tau} + N_{\bar{\tau}})
\end{equation}

Combining these four equations we can then extract 3 flavor fractions
($f_e\      =\ (N_e+N_{\bar{e}})/N_{total}$,
$f_{\mu}\  =\ (N_{\mu}+N_{\bar{\mu}})/N_{total}$,  and
$f_{\tau}\ =\ (N_{\tau}+N_{\bar{\tau}})/N_{total}$),
and the antiparticle to particle ratio, $N_{\bar{e}}/N_e$.

For simplicity we have ignored energy dependence and experimental factors of
acceptance, exposure, and efficiency.  A full error analysis is experimentally
dependent, and beyond the scope of this note.  However, we point out that by
employing ratios, particularly involving $N_1$, $N_3$, and $N_4$, much of the
systematic error will cancel.  In fact if we assume that we know
$N_{\bar{e}}/N_e$, then we can calculate the flavor ratios without employing
the muon measurements (which will have different systematic errors).  For the
more optimistic AGN fluxes cited\cite{Szabo94} the numbers of events
collected in a $km^3$ detector in one year would permit calculation of the
flavor content to a few percent.

\section{Atrophysical Neutrino Flavor Content}

In the absence of neutrino oscillations (discussed in the following section) we
expect almost no  $\nu_{\tau}$ content in astrophysical sources, as we discuss
in the following.

{}From the most discussed and seemingly most likely astrophysical high energy
neutrino sources we expect nearly equal numbers of particles and
anti-particles, half as many $\nu_e$'s as $\nu_{\mu}$'s and virtually no
$\nu_{\tau}$'s.  This comes about simply because the neutrinos are thought to
originate in decays of pions (and kaons) and subsequent decays of muons.  Most
astrophysical targets are tenuous even compared to the earth's atmosphere, and
would allow for full muon decay in flight.

The extreme case of an astrophyscial beam dump target of sufficient density to
suppress $\pi$ and $k$ decay (as would happen with a nucleon beam striking a
neutron star), but not so thick as to absorb the directly produced fluxes (a
neutron star would absorb almost any but a tangential netrino beam), seems to
be unlikely. Beyond the geometric difficulty, producing a detectable flux of
high energy neutrinos from such an inefficient source makes it even more
improbable.  There are some predictions for flavor independent fluxes from
cosmic defects and exotic objects such as evaportaing black holes.  Observation
of tau neutrinos from these would have great importance.  Indeed as we show
below, for such exotic sources along with the presence of oscillations, we
would have a unique flavor signature permitting unravelling the source content.

The flux ratio of $\nu_\mu : \nu_e : \ \nu_\tau = 2:1:0$ is  certainly valid
for AGN models such as the ones due Stecker  et al \cite{Stecker91} in which
the
neutrinos come from $\pi's$ produced  via the process $\gamma + p \rightarrow N
+ \pi$.  In the  calculations of Protheroe and Szabo [2] there are two
additional features:  the additional $\nu_e$ flux due to  escaping neutrons
modifies the $\nu_\mu : \nu_e$ flux ratio  to 1.75 : 1 and about 10\% of the
$\nu$-flux is due to proton-proton ($pp$) interactions.  Some fraction of the
$pp$ neutrinos will then give rise to neutrinos from rapidly decaying
secondaries containing heavy quarks (``prompt" production).

Depending on the amount of prompt $\nu$-flux  due to $D_S$ and $B \bar{B}$
production (and decay) there could be a  non-zero $\nu_\tau$ component present.
We now turn to a numerical estimate of such an initial $\nu_\tau$ flux
fraction.
The relative fluxes of prompt neutrino flavors can be written as
\cite{deRujula93}:

\[ \nu_\mu : \nu_e : \nu_\tau = 1 : 1 : t,  \]

where

\begin{equation}
t=\left \{
\frac{f}
 {B_\mu^D/B_\tau^s \ +
f (1 + 2f) B_\mu^s/
(2 + f) B_\tau^s}
+ \epsilon
\left ( \frac{B_\tau^b}{B_\mu^b} \right )  
\left ( \frac{B_\mu^b}{B_\mu^D} \right )
\right \}
/\left [ 1+ \epsilon
\frac{B^b_\mu}{ B_\mu^D} \right ]
\end{equation}

and where $f = \sigma_{D_{s}}/ \sigma_{D \bar{D}}, \ \epsilon \ = \sigma_{b
\bar{b}}/\sigma_D, B_\mu^D= \frac{1}{2} (B_\mu^{D^{0}}  + B_\mu^{D^{+}}) \ =
0.125, B^S_\tau \ = B r (D_s  \rightarrow \tau \nu_\tau) \sim 0.04, B^s_\mu =
Br (D_s \rightarrow \mu x) \ = 0.08, B_\tau^b =  0.025$ and $B_\mu^b = 0.10$
\cite{PDG92}.  Using a value for $f \sim  0.15$ \cite{Shipbaugh88} and
$\epsilon \sim 0.1$, we find the right hand side to be  $1 : 1 : (0.07 \sim
0.1)$.  Hence, in the Protheroe-Szabo  AGN model, the fluxes go as

\[ \nu_\mu : \nu_e : \nu_\tau \ = 1.0 : 0.6 : \ <0.01. \]

Actually this is a conservative upper bound on the $\nu_{\tau}$ content
because most of the secondary pions will have ample opportunity to decay and in
that case the $\tau$ fraction will be nearer $10^{-4}$.

Even in the extreme case that the matter density is so high that
only the prompt neutrinos survive (for which as stated earlier we know of no
proposed astrophysical situation), the flux ratios are:

\[ \nu_\mu : \nu_e : \nu_\tau \ = 1.0 : 1.0 : \ <0.1. \]

The conclusion is that the most likely expected astrophysical neutrino sources
will have very little $\nu_{\tau}$ flux, and that observation of such particles
is a signal for either or both of dramatic new sources and neutrino mass and
oscillations.

\section{Sensitivity to Neutrino Oscillations}

The $\delta m^2$ sensitivity (from $L/E$) for this observation is then
fantastic, going down to the
order of $10^{-16}\ eV^2$ (the distance is out to the AGN, $\sim 100\ Mpc$).
To determine the two neutrino mixing angle sensitivity limit requires detector
specific simulations.  The limitation has to do with the AGN neutrino flux
magnitude and effective volume for these events, and will probably be limited
by statistics, at least in the near future.  We guess it will be no better than
$0.01$ in the best of situations (as with most oscillation experiments).

To discuss the (to be observed) fluxes in terms of neutrino oscillations we
make a number of simplifying (but reasonable) assumptions.  Explicitly, we
assume that:

\begin{enumerate}

\item the initial fluxes are in the proportion
$\nu_{\mu}$ : $\nu_{e}$ : $\nu_{\tau}$ :: 2 : 1 : 0 (as generally expected),

\item there are equal numbers of neutrinos and anti-neutrinos (although this is
not crucial and can easily be dropped in actual analysis),

\item all the $\delta m^2$ are above $10^{-16}\ eV^2$, and that
$sin^2(\delta m^2 L/4 E)$ all average to $1/2$,

\item matter effects at the production site are negligible (this is reasonable
since $N_{e^-} \simeq N_{e^+}$ and baryon densities are low for AGN), and

\item  that there are no large matter effects in the path to earth.  The latter
is reasonable in the $\delta m^2$ range of interest ($10^{-2}$ to
$10^{-6}\ eV^2$).

\end{enumerate}

With these assumptions, let us first consider the case where the low energy
atmospheric neutrino anomaly is accounted for by simple two flavor neutrino
oscillations  \cite{atmanomaly}.  The oscillations could be either $\nu_{\mu}
\leftrightarrow \nu_{e}$ or $\nu_{\mu} \leftrightarrow \nu_{\tau}$ with $\delta
m^2$ of $10^{-2}-10^{-3}\ eV^2$ and $sin^2 2 \theta$ ranging from $0.5$ to
$1.0$.

In both cases the $\nu_{\mu}/\nu_e$ ratio should be modified {\bf exactly} as
in the atmospheric case\cite{atmanomaly}:  given that the $\nu_{\mu}/\nu_e$ is
found to be $0.6$ of the expected value of $2$, we expect exactly the same
thing from the distant cosmic sources of much higher energy, $\nu_{\mu}$ :
$\nu_e$ = 1.2 : 1. In the $\nu_{\mu} \leftrightarrow \nu_e$ mixing case this
translates into  $\nu_{\mu}$ : $\nu_{e}$ : $\nu_{\tau}$ :: 1.64 :  1.36 : 0.0
(using the earlier normalization to 3 total). In the other case, of $\nu_{\mu}
\leftrightarrow \nu_{\tau}$ mixing, we predict  $\nu_{\mu}$ : $\nu_{e}$ :
$\nu_{\tau}$ :: 1.2 : 1.0 : 0.8. It is easy to see that in the two flavor
mixing case, $\nu_e/\nu_{\mu}$ can never be greater than one.  Even when the
mixing is maximal on finds $\nu_{\mu}$ : $\nu_{e}$ : $\nu_{\tau}$ :: 1.5 : 1.5
: 0.0 in the $\nu_{\mu} \leftrightarrow \nu_e$ mixing case, and $\nu_{\mu}$ :
$\nu_{e}$ : $\nu_{\tau}$ :: 1 : 1 : 1 in the $\nu_{\mu} \leftrightarrow
\nu_{\tau}$ case.

We can now ask how expectations will change if we include neutrino oscillation
solutions to the solar neutrino deficit.  We can consider two distinct
possibilities:

\begin{enumerate}

\item The $\nu_e$ mixing is with either other species with a small
angle, as in the small angle MSW regime, then the above results remain
unaffected.

\item All three flavors mix substantially and the $\delta m^2$'s are in the
range of $10^{-5}$ to $10^{-6}\ eV^2$ (corresponding to the MSW ``large angle''
solution) or $10^{-10}\ eV^2$ (corresponding to the ``long wavelength'' case).

\end{enumerate}

The parameter space for a combined fit to the atmospheric and solar neutrino
data allowing for three flavor mixing has been given for each of the above
cases by Fogli, {\it et al.}\cite{Fogli93}, and by Acker, {\it et
al.}\cite{Acker93}, respectively.  We have evaluated the survival and
transition probabilities for the whole range of allowed values of the three
mixing angles, and the results are shown in Figure \ref{fig:FRACTIONS}.  This
figure plots the fraction of muon neutrinos versus the fraction of electron
neutrinos, and thus where each point specifies a fraction of tau neutrinos  (it
is analogous to the color triangle). The initial expected flux ($\nu_{\mu}$ :
$\nu_{e}$ : $\nu_{\tau}$ :: 2 : 1 : 0) is at $f_{\mu}\ =\ 0.66$ and $f_e\ =\
0.33$.

We further observe that:

\begin{enumerate}

\item Almost all combinations of acceptable mixing angles result in saturation
values to be observed with AGN neutrinos which lie between the lines
$f_e+f_{\mu}\ =\ 0.88$ and  $f_e+f_{\mu}\ =\ 0.66$.  Hence a substantial
number of $\nu_{\tau}$ events are expected ($0.34 > f_{\tau} > 0.12$) for all
situations which solve the solar and atmospheric problems with neutrino
oscillations.

\item Since in the case of two flavor mixing it is impossible to obtain
$f_e/f_{\mu} > 1$, observation of the data falling above the diagonal
$f_e = f_{\mu}$ line is clear evidence for three flavor mixing.

\end{enumerate}

We would like to stress that a small ``impurity" of  $\nu_\tau's$ in the
initial neutrino fluxes, as discussed in the previous section, has a negligible
effect on the $\nu_\tau$ signal due to  $\nu_\mu - \nu_\tau$ oscillations with
moderately large mixing angles.  For example, in the Protheroe-Szabo case,
(for $P_{\mu \mu} \sim 0.6$ and $P_{\mu \tau} \sim 0.4)$ the fluxes at the
earth become:

\[ \nu_\mu: \nu_e: \ \nu_\tau = 1.0 : 1.0 :  0.7, \]

which is quite similar to the pattern in the case above.

Interestingly, the purely prompt flux case, for which we are not aware of
any model, gives rise to a very distinct flux pattern on arrival,

\[ \nu_\mu: \nu_e : \nu_\tau \ = 1.0 : 1.56 : 0.72 = 0.3 : 0.47 : 0.23. \]

This represents a very singular point in Figure \ref{fig:FRACTIONS},
distinguished by the dominance of the $\nu_e$ flux.  We also note in passing
that in the unexpected case of initial flavor independent flux, the equal
fluxes remain so in any combination of neutrino mass and mixing. We thus
conclude that almost any astrophysical source of neutrinos will result in
significant numbers of $\nu_{\tau}$'s, if neutrinos have mass and they
oscillate.

\section{Backgrounds: Almost None}

By background we mean events which can fake the double bang signature of two
huge cascades spaced roughly $100~m$ apart, with the second larger than the
first (by typically a factor of two), and the sum of the visible energy being
in the range of a few $PeV$.  There are two general types of possible
backgrounds due to neutrinos: those into which the neutrino transforms ($e$,
$\mu$ and $\tau$), and particles generated in the cascade resulting from the
energy transferred to the struck quark.  Neutral current events with an
outgoing neutrino have interaction lengths so great as to be negligibly likely
to interact again at close range ($<10^{-6}$ at this $PeV$ energy).  Electrons
will immediately radiate, and be added to the initial a cascade, and be
indistinguishable from neutral current events.

The most serious background is due to $\nu_\mu \ cc$ events in  which the muon
travels about 100 m and then loses most of  its energy in a single catastrophic
Bremsstrahlung radiation.  Such events will have energy characteristics
similar to signal events.  The rate for such events  as a fraction of $cc$
events can be estimated as

\begin{equation}
\left ( \frac{m_e}{m_\mu} \right )^2   \ \left (
\frac{100m}{R} \right ) \
\left ( \frac{\Delta E}{E} \right )
\end{equation}

where $R$ is the radiation length in water and $\Delta E/E  \sim 0 (1/3)$ for
the fraction of energy deposited.  This number is of the order $3\cdot 10^{-3}$
and is reassuringly small although a more accurate calculation is called for.

Once again, detector dependent simulations are required to numerically evaluate
the confusion probability.  Whatever that probability is, and we expect it to
be low at $2\ PeV$, it decreases rapidly with increasing energy (short range
muon radiation goes up while the tau decay length increases).  In a large
array, one may also demand that there be no continuing track, since the
probablility of a $2~PeV$ muon stopping after $100~m$ is negligible.  However,
we estimate that this constraint will not be needed.

The other potential source of background from neutrino interactions is from the
hadronic side of the neutrino interaction vertex.  First of all, we do not know
of any particle with the ability to penetrate $\sim 10^4\ gm/cm^2$ of matter
before decaying or interacting  ($100\ m$ is about $100$ strong interaction
lengths in water), except for  leptons.  Thus the only situation with which we
must be concerned is when a $\tau$ results from the decays of particles in the
cascade, and moreover when this $\tau$ has most of the energy, which we
show below to be quite unlikely.

One source of this type of background is the production (and decay to
$\tau$) of $D_s$ (and $B \bar{B}$ pairs) in neutral current  interactions of
$\nu_e$ (and $\nu_\mu)$ and charged current  interactions of $\nu_e$.  These
could give rise to ``double bang"  events like the $\nu_\tau$ events if $D_S$
(or B) should  decay within 10 m to produce an energetic $\tau$ which then
travels  about 100 m before decaying.  In such events the average energy of
the second bang $E_2 \sim 2/3 E_\tau \sim 1/2 E_h \sim 1/8 z E_\nu$ where
$z=E_h/E_\nu$. Since $z$ is expected to be  rather small, we expect $E_2/E_1 <<
1$.  Hence the  $E_2/E_1 >  1$ cut will remove most such events. The number of
such events will be $N \sim \left \{ rN_\mu +  (1+r) N_e \right \}  (f_{D_{s}}
B_\tau^s + f^2_B  B^B_\tau)$ where$f_i$ are the fractional production of $D_s$
and $B \bar{B}$ respectively.  For $f_i \sim 10^{-2}, r \sim  1/3, N_\mu =
N_e$; we find $N \sim 1.5\cdot 10^{-3} N_\mu$, most of which will be removed by
the
energy fraction cut.

A somewhat more serious background is diffractive production  of $D_s$ via
charm changing $cc$ interaction of $\nu_e$ in  the reaction $\nu_e + N
\rightarrow e + D_s + x.$ There is no Cabibbo suppression, the rate can be as
high as  $6\cdot 10^{-3}$ times the $cc$ rate and furthermore $D_s$ can  have
average energy of about 1/2 $E_\nu$ [13].  Thus  $E_2/E_1$ can be of order 1
and some events may pass the  $E_2/E_1$ cut.  However, the rate is given by
$6\cdot  10^{-3}\cdot B_\tau^{Ds}  \sim 2.5\cdot 10^{-4} N_e$ and is thus also
quite small.

Another class of potential background might be due to downgoing muons which
originate in primary cosmic ray interactions in the atmosphere.  The angular
distribution of the downgoing muons peaks strongly near the zenith, while the
double bang events would be uniform in direction in the upper hemisphere.  A
stronger constraint is that events with such high energy are not expected from
downgoing muons at depths of a few $km$.  Bremsstrahlung and pair production
events of a few hundred $GeV$ are plentiful, but the flux falls very fast with
energy, and negligible numbers are expected above  $100\ TeV$\cite{Okada94}.
The probability of observing two cascades meeting the spacing, total energy and
energy sharing requirement is vanishingly small.

There are also muons arriving from all directions, due to distant neutrino
interactions.  Clearly if the muon charged current interaction is not a
problem, as discussed above, then the probability of two bursts of adequate
energy and spacing will not be a problem, even without invoking observation of
incoming or outgoing muons as a veto.

Finally, we note that as a check, studies of the decay path length distribution
requiring consistency with the known tau lifetime, should confirm the detection
of taus and the lack of contamination of the sample.  In sum we conclude that
the data selection criteria on spacing and energy of the two bangs will remove
virtually all backgrounds and leave a verifiably pure sample of UHE $\tau$
events.

\section{Old Idea, New Combination and Impact}

The above is not entirely a new idea. Several authors have written about $\tau$
signatures in the past\cite{Learned80}. The new recognition is in coupling the
uniqueness of the double bang signature, the encouraging recent flux
predictions from AGN, plus the modern hints about neutrino oscillations, and
thus being able to claim to have a good chance to observe a pure sample of
$\nu_{\tau}$ interactions in detectors within the realm of possibility in the
next few years.  Further we now recognize that detection of this novel class of
interactions makes possible observations of the neutrino flavor mix,
oscillations, distinction of some exotic class of sources, precise measurements
of the neutrino cross section in the PeV energy range, and more distantly in
the future, earth tomography with greater sensitivity than has been thought
possible.

One impact of the consideration of this potential observation gives motivation
to fill the volume, somewhat, of a hypothetical  $1\ km^3$ array.  Indeed this
experiment itself could perhaps justify construction of that instrument.

Also, given that these events are near the anticipated acoustic detection
threshold\cite{Learned93}, one may contemplate hearing the double clicks from
such events at few $km$ ranges and higher energies.  The unusual property of
acoustic pulse amplitude which results in slow decrease with distance in water
may make practical the search for higher energy neutrino interactions from
volumes of many cubic kilometers.  Additionally, the possibility of detecting
neutrino induced cacades via radio pulses of $GHz$ frequency range continues to
be explored in deep ice\cite{Smoot94}.  As for acoustic sensing, the double
microwave pulses due to $\tau$ production and decay should provide dramatic and
unique signatures.

We remind the reader that all of the above requires the existence of
substantial numbers of $PeV$ neutrinos, which matter should be resolved in the
next few years by the AMANDA, Baikal, DUMAND, and NESTOR experiments now under
construction. If those ultra high energy neutrinos are present in expected
numbers, then we believe that the observation of the double bang events, along
with other previously discussed interactions, will lead to important particle
physics measurements which cannot be carried out in any other way on earth.

\section*{Acknowledgement}

We want to acknowledge discussions with F.~Halzen, M.~Jones, L.~Resvanis,
J.~Sender, V.~Stenger, H.~Sugawara, S.~F.~Tuan, K.~Winter, and especially
helpful and clarifying discussions with J.~Bjorken and X.~Tata. This work was
supported in part by U.S. DOE Grant DE-FG-03  94ER40833.


\newpage

\begin{figure}
\setlength{\unitlength}{0.012500in}%
\begingroup\makeatletter\ifx\SetFigFont\undefined
\def\x#1#2#3#4#5#6#7\relax{\def\x{#1#2#3#4#5#6}}%
\expandafter\x\fmtname xxxxxx\relax \def\y{splain}%
\ifx\x\y   
\gdef\SetFigFont#1#2#3{%
  \ifnum #1<17\tiny\else \ifnum #1<20\small\else
  \ifnum #1<24\normalsize\else \ifnum #1<29\large\else
  \ifnum #1<34\Large\else \ifnum #1<41\LARGE\else
     \huge\fi\fi\fi\fi\fi\fi
  \csname #3\endcsname}%
\else
\gdef\SetFigFont#1#2#3{\begingroup
  \count@#1\relax \ifnum 25<\count@\count@25\fi
  \def\x{\endgroup\@setsize\SetFigFont{#2pt}}%
  \expandafter\x
    \csname \romannumeral\the\count@ pt\expandafter\endcsname
    \csname @\romannumeral\the\count@ pt\endcsname
  \csname #3\endcsname}%
\fi
\fi\endgroup
\begin{picture}(422,325)(78,440)
\thinlines
\put(141,699){\circle*{6}}
\put(141,657){\circle*{6}}
\put(141,614){\circle*{6}}
\put(141,572){\circle*{6}}
\put(141,530){\circle*{6}}
\put(141,488){\circle*{6}}
\put(141,445){\circle*{6}}
\put(310,741){\circle*{6}}
\put(310,699){\circle*{6}}
\put(310,657){\circle*{6}}
\put(310,614){\circle*{6}}
\put(310,572){\circle*{6}}
\put(310,530){\circle*{6}}
\put(310,488){\circle*{6}}
\put(310,445){\circle*{6}}
\put(479,741){\circle*{6}}
\put(479,699){\circle*{6}}
\put(479,657){\circle*{6}}
\put(479,614){\circle*{6}}
\put(479,572){\circle*{6}}
\put(479,530){\circle*{6}}
\put(479,488){\circle*{6}}
\put(479,445){\circle*{6}}
\put(141,572){\circle{22}}
\put(141,614){\circle{10}}
\put(141,530){\circle{32}}
\put(141,488){\circle{22}}
\put(141,445){\circle{10}}
\put(310,614){\circle{10}}
\put(141,741){\circle*{6}}
\put(310,657){\circle{10}}
\put(109,464){\makebox(0,0)[lb]{\smash{\SetFigFont{12}{14.4}{rm}Neutrino
Interaction}}}
\put(310,572){\circle{12}}
\put(141,657){\circle{10}}
\put(479,699){\circle{42}}
\put(479,741){\circle{32}}
\put(479,657){\circle{32}}
\put(479,614){\circle{22}}
\put(479,572){\circle{12}}
\thicklines
\put( 78,509){\line( 2, 1){392}}
\thinlines
\put(141,762){\line( 0,-1){317}}
\put(310,765){\line( 0,-1){320}}
\put(479,762){\line( 0,-1){317}}
\put(143,607){\vector( 1, 2){0}}
\multiput(
99,519)(3.98182,7.96364){11}{\makebox(0.1111,0.7778){\SetFigFont{5}{6}{rm}.}}
\put(138,567){\vector( 1, 2){0}}
\multiput(120,530)(3.68000,7.36000){5}{\makebox(0.1111,0.7778){\SetFigFont{5}{6}{rm}.}}
\put(138,530){\vector( 4,-1){0}}
\multiput(130,532)(8.00000,-2.00000){1}{\makebox(0.1111,0.7778){\SetFigFont{5}{6}{rm}.}}
\put(136,490){\vector( 3,-4){0}}
\multiput(110,525)(5.23200,-6.97600){5}{\makebox(0.1111,0.7778){\SetFigFont{5}{6}{rm}.}}
\put(306,652){\vector( 2, 3){0}}
\multiput(275,605)(5.17948,7.76922){6}{\makebox(0.1111,0.7778){\SetFigFont{5}{6}{rm}.}}
\put(305,611){\vector( 4,-1){0}}
\multiput(290,615)(7.41175,-1.85294){2}{\makebox(0.1111,0.7778){\SetFigFont{5}{6}{rm}.}}
\put(305,575){\vector( 4,-1){0}}
\multiput(245,590)(8.57143,-2.14286){7}{\makebox(0.1111,0.7778){\SetFigFont{5}{6}{rm}.}}
\put(479,738){\vector( 1, 2){0}}
\multiput(460,700)(3.80000,7.60000){5}{\makebox(0.1111,0.7778){\SetFigFont{5}{6}{rm}.}}
\put(478,662){\vector( 1,-1){0}}
\multiput(445,695)(6.50000,-6.50000){5}{\makebox(0.1111,0.7778){\SetFigFont{5}{6}{rm}.}}
\put(475,619){\vector( 1,-2){0}}
\multiput(440,690)(3.93333,-7.86667){9}{\makebox(0.1111,0.7778){\SetFigFont{5}{6}{rm}.}}
\put(101,469){\line(-1, 0){ 15}}
\put( 86,469){\vector( 1, 3){ 14.100}}
\put(197,538){\line(-1, 0){ 11}}
\put(186,538){\vector( 1, 2){  8.800}}
\put(411,725){\line( 1, 0){ 18}}
\put(429,725){\vector( 1,-1){ 21}}
\thicklines
\put( 80,510){\line( 6, 5){ 30}}
\put(110,535){\line( 1, 0){  5}}
\multiput(115,535)(0.25000,-0.50000){21}{\makebox(0.4444,0.6667){\SetFigFont{7}{8.4}{rm}.}}
\put(120,525){\line( 0,-1){  5}}
\put(120,520){\line(-4,-1){ 40}}
\put( 80,510){\line( 0, 1){  0}}
\put(430,685){\line( 6, 5){ 30}}
\put(460,710){\line( 1, 0){  5}}
\multiput(465,710)(0.25000,-0.50000){21}{\makebox(0.4444,0.6667){\SetFigFont{7}{8.4}{rm}.}}
\put(470,700){\line( 0,-1){  5}}
\put(470,695){\line(-4,-1){ 40}}
\put(430,685){\line( 0, 1){  0}}
\put(344,720){\makebox(0,0)[lb]{\smash{\SetFigFont{12}{14.4}{rm}Tau Decay}}}
\put(204,532){\makebox(0,0)[lb]{\smash{\SetFigFont{12}{14.4}{rm}Tau Track}}}
\end{picture}

\caption{A schematic view of a ``double bang'' event near a deep ocean
detector whose modules are indicated by dots.  Such cascades should be visible
to detectors from hundreds of meters distance.}
\label{fig:DOUBLEBANG}
\end{figure}

\begin{figure}

\setlength{\unitlength}{0.012500in}%
\begingroup\makeatletter\ifx\SetFigFont\undefined
\def\x#1#2#3#4#5#6#7\relax{\def\x{#1#2#3#4#5#6}}%
\expandafter\x\fmtname xxxxxx\relax \def\y{splain}%
\ifx\x\y   
\gdef\SetFigFont#1#2#3{%
  \ifnum #1<17\tiny\else \ifnum #1<20\small\else
  \ifnum #1<24\normalsize\else \ifnum #1<29\large\else
  \ifnum #1<34\Large\else \ifnum #1<41\LARGE\else
     \huge\fi\fi\fi\fi\fi\fi
  \csname #3\endcsname}%
\else
\gdef\SetFigFont#1#2#3{\begingroup
  \count@#1\relax \ifnum 25<\count@\count@25\fi
  \def\x{\endgroup\@setsize\SetFigFont{#2pt}}%
  \expandafter\x
    \csname \romannumeral\the\count@ pt\expandafter\endcsname
    \csname @\romannumeral\the\count@ pt\endcsname
  \csname #3\endcsname}%
\fi
\fi\endgroup
\begin{picture}(300,312)(20,500)
\thicklines
\put( 80,560){\line( 1, 1){240}}
\put( 80,680){\line( 1,-1){120}}
\multiput( 80,560)(12.61558,25.23117){10}{\line( 1, 2){  6.460}}
\put( 75,540){\makebox(0,0)[lb]{\smash{\SetFigFont{12}{14.4}{rm}0.0}}}
\put(315,540){\makebox(0,0)[lb]{\smash{\SetFigFont{12}{14.4}{rm}4.0}}}
\put( 50,560){\makebox(0,0)[lb]{\smash{\SetFigFont{12}{14.4}{rm}0.0}}}
\put( 50,795){\makebox(0,0)[lb]{\smash{\SetFigFont{12}{14.4}{rm}4.0}}}
\put(190,520){\makebox(0,0)[lb]{\smash{\SetFigFont{12}{14.4}{rm}E1}}}
\put( 80,560){\framebox(240,240){}}
\put(180,500){\makebox(0,0)[lb]{\smash{\SetFigFont{12}{14.4}{rm}( PeV)}}}
\put(160,740){\makebox(0,0)[lb]{\smash{\SetFigFont{12}{14.4}{rm}Region}}}
\put( 20,680){\makebox(0,0)[lb]{\smash{\SetFigFont{12}{14.4}{rm}( PeV)}}}
\put( 25,700){\makebox(0,0)[lb]{\smash{\SetFigFont{12}{14.4}{rm}E2}}}
\put(220,600){\makebox(0,0)[lb]{\smash{\SetFigFont{12}{14.4}{rm}Background }}}
\put(235,580){\makebox(0,0)[lb]{\smash{\SetFigFont{12}{14.4}{rm}Region}}}
\put(175,660){\makebox(0,0)[lb]{\smash{
\SetFigFont{12}{14.4}{rm}Require $E_2
> E_1$
}}}
\put( 95,670){\makebox(0,0)[lb]{\smash{
\SetFigFont{12}{14.4}{rm}Require $E_1
+ E_2 >E_{min}$
}}}
\put(170,765){\makebox(0,0)[lb]{\smash{\SetFigFont{12}{14.4}{rm}Signal}}}
\end{picture}

\caption{Diagramatic representation of the region of double bang signals and
background signals, as discussed in the text.  The signal is essentially
background free after cuts on minimum total energy and requiring the second
cascade to be the larger.} \label{fig:e1_vs_e2}
\end{figure}

\begin{figure}
\setlength{\unitlength}{0.012500in}%
\begingroup\makeatletter\ifx\SetFigFont\undefined
\def\x#1#2#3#4#5#6#7\relax{\def\x{#1#2#3#4#5#6}}%
\expandafter\x\fmtname xxxxxx\relax \def\y{splain}%
\ifx\x\y   
\gdef\SetFigFont#1#2#3{%
  \ifnum #1<17\tiny\else \ifnum #1<20\small\else
  \ifnum #1<24\normalsize\else \ifnum #1<29\large\else
  \ifnum #1<34\Large\else \ifnum #1<41\LARGE\else
     \huge\fi\fi\fi\fi\fi\fi
  \csname #3\endcsname}%
\else
\gdef\SetFigFont#1#2#3{\begingroup
  \count@#1\relax \ifnum 25<\count@\count@25\fi
  \def\x{\endgroup\@setsize\SetFigFont{#2pt}}%
  \expandafter\x
    \csname \romannumeral\the\count@ pt\expandafter\endcsname
    \csname @\romannumeral\the\count@ pt\endcsname
  \csname #3\endcsname}%
\fi
\fi\endgroup
\begin{picture}(372,378)(8,447)
\thinlines
\put(271,613){\circle{8}}
\put(219,665){\circle{8}}
\put(155,660){\circle{10}}
\put( 61,823){\line( 1,-1){317}}
\put( 61,506){\line( 1, 1){158.500}}
\thicklines
\put(168,613){\line( 1, 1){ 51.500}}
\thinlines
\put(287,641){\vector(-2,-3){ 12.923}}
\put(211,538){\vector(-1, 2){ 15.800}}
\put(184,649){\line(-1,-1){ 24}}
\put(160,625){\line( 0,-1){ 16}}
\put(160,609){\line( 1,-1){ 27.500}}
\put(188,582){\line( 3,-1){ 83.100}}
\put(271,554){\line( 0, 1){ 20}}
\put(271,574){\line(-5, 4){ 43.902}}
\put(227,609){\line(-3, 1){ 38.700}}
\put(188,621){\line( 2, 3){ 10.769}}
\put(199,637){\line(-5, 4){ 15}}
\put(184,649){\line( 0, 1){  0}}
\put(188,613){\line(-1, 1){ 20}}
\multiput(168,633)(-0.40000,-0.40000){21}{\makebox(0.1111,0.7778){\SetFigFont{5}{6}{rm}.}}
\put(160,625){\line( 0,-1){ 16}}
\put(160,609){\line( 4, 3){ 16}}
\multiput(176,621)(0.48000,-0.32000){26}{\makebox(0.1111,0.7778){\SetFigFont{5}{6}{rm}.}}
\put(188,613){\line( 0, 1){  0}}
\put(260,690){\vector(-1,-1){ 19.500}}
\put(191,666){\vector(-1,-1){ 16}}
\thicklines
\put( 60,823){\line( 0,-1){317}}
\put( 60,506){\line( 1, 0){317}}
\thinlines
\multiput(140,585)(-0.41667,0.41667){13}{\makebox(0.1111,0.7778){\SetFigFont{5}{6}{rm}.}}
\put( 60,665){\line(-1, 0){  5}}
\put( 60,825){\line(-1, 0){  5}}
\put( 60,505){\line(-1, 0){  5}}
\put( 60,505){\line( 0,-1){  5}}
\put(168,613){\circle{8}}
\put(220,505){\line( 0,-1){  5}}
\put( 45,500){\makebox(0,0)[lb]{\smash{
\SetFigFont{14}{16.8}{rm}0.0
}}}
\put(380,505){\line( 0,-1){  5}}
\put(115,695){\vector( 4,-3){ 31.200}}
\put(156,447){\makebox(0,0)[lb]{\smash{\SetFigFont{14}{16.8}{rm}Muon
Fraction}}}
\put( 25,601){\makebox(0,0)[lb]{\smash{
\SetFigFont{14}{16.8}{rm}Electron
Fraction
}}}
\put(295,657){\makebox(0,0)[lb]{\smash{\SetFigFont{14}{16.8}{rm}Expected}}}
\put(295,637){\makebox(0,0)[lb]{\smash{\SetFigFont{14}{16.8}{rm}Cosmic}}}
\put( 57,479){\makebox(0,0)[lb]{\smash{\SetFigFont{14}{16.8}{rm}0.0}}}
\put(370,479){\makebox(0,0)[lb]{\smash{\SetFigFont{14}{16.8}{rm}1.0}}}
\put(207,665){\makebox(0,0)[lb]{\smash{
\SetFigFont{14}{16.8}{rm}0.0
}}}
\put(128,586){\makebox(0,0)[lb]{\smash{
\SetFigFont{14}{16.8}{rm}0.5
}}}
\put(295,617){\makebox(0,0)[lb]{\smash{\SetFigFont{14}{16.8}{rm}no Osc}}}
\put(219,538){\makebox(0,0)[lb]{\smash{\SetFigFont{14}{16.8}{rm}3 Neutrino
Mix}}}
\put(219,517){\makebox(0,0)[lb]{\smash{\SetFigFont{14}{16.8}{rm}Allowed
Region}}}
\put(265,695){\makebox(0,0)[lb]{\smash{\SetFigFont{14}{16.8}{rm}Underground}}}
\put(265,720){\makebox(0,0)[lb]{\smash{\SetFigFont{14}{16.8}{rm}Observed }}}
\put( 70,700){\makebox(0,0)[lb]{\smash{\SetFigFont{14}{16.8}{rm}Prompt Nus}}}
\put( 90,575){\makebox(0,0)[lb]{\smash{
\SetFigFont{14}{16.8}{rm}Tau
Fraction
}}}
\put( 45,815){\makebox(0,0)[lb]{\smash{
\SetFigFont{14}{16.8}{rm}1.0
}}}
\end{picture}

\caption{The fraction of muon neutrinos versus electron neutrinos, allowing for
a fraction of tau neutrinos.  Expected initial flux is at 2/3, 1/3. Full mixing
would result in 1/3, 1/3.  The points represent various solutions to the solar
and atmospheric neutrino problems.  The point corresponding to pure prompt
$\nu$-beam is at 0.30, 0.47 \& 0.23 taus.} \label{fig:FRACTIONS}
\end{figure}


\begin{thebibliography}{99}

\bibitem{DUMAND} We use DUMAND herein to refer to a generic deep ($km$ or more)
under water or under ice detector which senses Cherenkov radiation with
photomultiplier tubes at distances of tens of meters from the tracks.  Arrays
with dimensions of order $100~m$ are now under construction in the ocean
(DUMAND and NESTOR), a lake (Baikal) and the polar ice (AMANDA).  See
review paper by J.~G.~Learned, Phil. Trans. R. Soc. Lond. A {\bf 346}, 99
(1994), and references therein.  A future $km$ scale detector is presently
under discussion in the community and could be constructed within a decade.

\bibitem{Stenger92}  ``Proceedings of the Workshop  on High Energy Neutrino
Astrophysics'', 23-26 March 1992, V.~J.~Stenger, {\it et al.}, eds.,
World Scientific
(1992).

\bibitem{PDG92} M. Aguilar-Benitez, {\it et al.}, ``Review of Particle
Properties'', Phys. Rev. D {\bf 45}, Part 2 (1992).

\bibitem{Szabo94} A.~P.~Szabo and R.~J.~Protheroe, Adelaide preprint
ADP-AT-94-4, to be published in Astroparticle Physics
(1994), Ref. 1, p. 24.

\bibitem{Quigg86} C.~Quigg, M.~H.~Reno, and T.~P.~Walker, Phys. Rev. Let.
{\bf 57}, 774 (1986).

\bibitem{Stecker91} F.~Stecker {\it et al.,} Phys. Rev. Lett.
{\bf 66}, 2697 (1991), Ref. 1, p. 1.

\bibitem{deRujula93} See for example, A. de Rujula, F. Fernandez and
J. J. Gomez-Cadenas, Nucl. Phys. {\bf B405}, 80 (1993).

\bibitem{Shipbaugh88} C.~Shipbaugh {\it et al.,} Phys. Rev. Lett. {\bf
60}, 2117 (1988).

\bibitem{atmanomaly} See Y.~Fukuda, {\it et al.}, ICRR-Report-321-94-16,
submitted to Phys. Lett. B, June 1994.  Also for a survey of the
$\nu_{\mu}/\nu_e$ anomaly, see the various papers from IMB, Kamioka and Soudan
submitted to the Neutrino '94 Conference, Eilat, June 1994, A. Dar, ed. in
press.

\bibitem{Fogli93} G.~L.~Fogli, E.~Lisi, and D.~Montanino,
Phys. Rev. {\bf D49}, 3626 (1994).

\bibitem{Acker93} A.~Acker, A.~B.~Balantekin, and F.~Loreti, Phys. Rev. D,
{\bf 49}, 328 (1994).

\bibitem{13} A.~E.~Asratyan {\it et al.,} Z. fur Phys. {\bf C58}, 55 (1993).

\bibitem{Okada94} A.~Okada, ICRR preprint, March 1994.

\bibitem{Learned80} J.~G.~Learned, ``Proceedings of the 1980 International
DUMAND Symposium'', V.~J.~Stenger, ed., {\bf II}, (1980), p. 272, and
references therein.

\bibitem{Learned93} J.~G.~Learned and R.~J.~Wilkes, 23 ICRC, Calgary (1993),
and B.~Price, Nuc. Inst. and Meth. {\bf A325}, 346 (1993).

\bibitem{Smoot94} New calculations and efforts at measuring background noise
and attenuation were reported by John Ralston and by George Smoot and
collaborators at the Snowmass '94 Workshop. These efforts may lead to radio
detection with an energy threshold lower than a $PeV$.

\end{thebibliography}
\end{document}